\documentclass[aps,prl,twocolumn,showpacs,preprintnumbers,amsmath,amssymb,
epsf,superscriptaddress,groupedaddress, tightenlines]{revtex4}

\usepackage{graphicx}
\usepackage{amsmath}
\usepackage{bm}
\newcommand{\nn}{\nonumber}

\def\OMIT#1{}

\begin{document}
\preprint{MIT-CTP 3723}

\title{Occupation number formalism for arbitrary $N_c$ baryons}

\author{Dan Pirjol}
\affiliation{Center for Theoretical Physics, Massachusetts Institute for
  Technology, Cambridge, MA 02139}

\author{Carlos Schat}
\affiliation{CONICET and Departamento de F\'{\i}sica, FCEyN, Universidad de Buenos
Aires, Ciudad Universitaria, Pab.1, (1428) Buenos Aires,
Argentina}

\date{\today}

\begin{abstract}
A general method is presented for computing matrix elements of
quark operators on baryonic states with low strangeness
and arbitrary number of colors $N_c$. These results are useful in
applications of the large $N_c$ expansion to baryons and exotics.
As an application we compute the matrix elements of strangeness
changing operators contributing to kaon couplings to ground state
baryons and pentaquarks, in broken SU(3).
\end{abstract}


\maketitle

1. \textit{Introduction.}
In the large $N_c$ limit, with $N_c$ the number of colors, a new symmetry
emerges in the baryon sector of QCD. This is the contracted symmetry
$SU(2F)_c$, with $F$ the number of light quark flavors \cite{Largenspinflavor,DJM}.
This symmetry can be used to organize the $1/N_c$ expansion of any
quantity as an operator expansion.
The nonrelativistic quark model is a convenient bookkeeping tool for
implementing this expansion \cite{largen,heavybaryons}.
The baryons are constructed by placing the $N_c$ quarks
into one-body states. For example, taking $F=3$ a possible
basis of one-body states consists of
\begin{eqnarray}\label{basis}
|u_\uparrow\rangle \,, |u_\downarrow\rangle \,,
|d_\uparrow\rangle \,, |d_\downarrow\rangle \,,
|s_\uparrow\rangle \,, |s_\downarrow\rangle \ ,
\end{eqnarray}
transforming in the fundamental representation of SU(6).
The operators representing physical quantities such as
masses, axial currents, etc. can be constructed from quark operators
annihilating the basis states in Eq.~(\ref{basis}). The
building blocks are bilinears of the form $q^\dagger \Lambda^A q$
with $\Lambda^A$ the generators of SU(2F)
\begin{eqnarray}\label{generators}
 J^i = q^\dagger (\frac{\sigma^i}{2} \otimes 1 ) q&,& \nonumber \\
 T^a = q^\dagger (1 \otimes \frac{\lambda^a}{2} ) q&,& \\
 G^{ia} = q^\dagger (\frac{\sigma^i}{2} \otimes \frac{\lambda^a}{2} ) q&.& \nonumber
\end{eqnarray}

One important technical problem in the implementation of this program
is the computation of the matrix elements of these operators on quark
model states. Various methods have been discussed in
the literature for this purpose. The computation
of these matrix elements for $F\geq 3$ {\em and} arbitrary $N_c$
turns out to be rather involved. For $F=2$ a method for computing
with any $N_c$ was discussed in Refs.~\cite{KaPa,Soldate,PY1}.
On the other hand, keeping $N_c=3$ (or other small values)
other methods are available, such as the holomorphic representation
of the harmonic oscillator discussed in \cite{hol}.
The difficulty is connected with the necessity of manipulating
complicated expressions involving SU(3) Clebsch-Gordan (CG) coefficients.
Although not insurmontable (applications of the large $N_c$
expansion in SU(3) have been presented in Refs.~\cite{SU3} and analytic expressions
for some arbitrary $N_c$ CG coefficients have been computed recently in Refs.~\cite{Clebsch}),
extracting the form of the result for arbitrary $N_c$ remains
a challenge.

In practice, we are interested only in baryon states with at most
a few strange or heavy quarks. Therefore it is natural to expect that the
large $N_c$ dependence comes from the large number of light quarks in the
$u,d$ sector.
In this paper we give the details required
for constructing explicitly any such states and computing matrix elements between them.
The advantage of our method is that only usual SU(2) CG coefficients
coefficients are ever required, and closed form expressions are found for
all matrix elements for arbitrary $N_c$.
\vspace{0.5cm}

2. \textit{States.} Consider the spin-flavor group SU(6), which contains
as a subgroup (see, e.g. Ref.~\cite{Georgi})
\begin{eqnarray}\label{decomp}
SU(6)_{SF} \supset SU(4)_{SI} \otimes SU(2)_K \otimes U(1)_{n_s} \ .
\end{eqnarray}
The factors on the right-hand side correspond to spin-isospin SU(4),
the strange quarks' spin SU(2)$_K$, and the U(1) associated with
strangeness $-n_s$.

We would like to construct the decomposition of a baryon state containing
$N_c$ quarks and transforming in the completely symmetric representation
$S_{N_c}$ of spin-flavor $SU(6)$ into irreducible representations of the
subgroup on the rhs of Eq.~(\ref{decomp}). This is given by
\begin{eqnarray}\label{SNcdecomp}
S_{N_c} = (S_{N_c}, {\mathbf 1}, 0) \oplus (S_{N_c-1}, {\mathbf 2}, 1)
\oplus (S_{N_c-2}, {\mathbf 3}, 2) \oplus \cdots
\end{eqnarray}
We denoted the representations of SU(2)$_K$ by their multiplicity
$2K+1$, with $K$ the spin of the strange quarks. The terms written
have $K=|n_s|/2$, which corresponds to the maximally possible value of
the strange quark spin. This is required by Fermi statistics as applied to the
system of the strange quarks, assuming that they are all in a
completely symmetric orbital wave function. This is satisfied by
ground state baryons, but not by orbitally excited states. We will comment
on this case below.

The wave function of a hadron containing $N_c$ quarks, of which $n_s$ are strange
quarks, factors according to Eq.~(\ref{SNcdecomp}) into a product of
wave functions for its components. The nonstrange system has
a symmetric spin-flavor wave function. For describing its state, it is
convenient to use a Fock state formalism, familiar from the theory of
many-body systems~\cite{manybody}.

Since the orbital wave functions of all quarks are the same,
we can label the state of a system of $N_c$ identical quarks by
giving the occupation numbers of the one-body spin-flavor
states in Eq.~(\ref{basis}).
We introduce the ``6n-symbol'' defined as
\begin{eqnarray}\label{quark}
&& \{ n_1, n_2, n_3, n_4, n_5, n_6 \} =
\sqrt{\frac{n_1 ! n_2 ! n_3 ! n_4 ! n_5! n_6!}{N !}} \nn \\
&&
\times (u_\uparrow^{n_1}u_\downarrow^{n_2}d_\uparrow^{n_3}d_\downarrow^{n_4}
s_\uparrow^{n_5}s_\downarrow^{n_6} + \mbox{perms})\, ,
\end{eqnarray}
with $N = \sum_{i=1}^6 n_i$.
These states are normalized as
\begin{eqnarray}
&& \langle \{ n'_1, n'_2, n'_3, n'_4, n'_5, n'_6 \} | \{ n_1, n_2, n_3, n_4,
n_5, n_6 \}
\rangle  \nn \\
&& \qquad = \delta_{n_1 n'_1} \delta_{n_2 n'_2}\delta_{n_3 n'_3}\delta_{n_4 n'_4}
\delta_{n_5 n'_5}\delta_{n_6 n'_6}\,.
\end{eqnarray}
Nonstrange hadrons have $n_5=n_6=0$ and can be described by ``4n-symbols''
$\{n_1, n_2, n_3, n_4 \}$. For simplicity we will use this notation when
appropriate.

The nonstrange states have spin and
isopin satisfying $I=J$. Their spin-flavor symmetric wave functions can be given in closed form
as \cite{Soldate,PY1}
\begin{eqnarray}\label{nons}
&& |I I_3 J_3 ; N_{ud} \rangle = \sum_i
\left(
\begin{array}{cc|c}
\frac{N_u}{2}& \frac{N_d}{2} & I \\
i & J_3-i & J_3 \\
\end{array}
\right) \\
&& \quad \times \{ \frac{N_u}{2}+i, \frac{N_u}{2}-i,
\frac{N_d}{2}+J_3-i,
\frac{N_d}{2}-J_3+i\} \ , \nn
\end{eqnarray}
where $N_{u,d}$ are the number of up and down quarks, respectively, and $N_{ud}=N_c-n_s$.
\begin{eqnarray}
N_u = \frac{N_{ud}}{2}+I_3\,,\quad
N_d = \frac{N_{ud}}{2}-I_3 \ .
\end{eqnarray}
A few representative nonstrange $J_3 = +\frac12$ states are
\begin{eqnarray}
p_\uparrow &=& \sqrt{\frac23} \{2,0,0,1\} - \frac{1}{\sqrt3}
\{1,1,1,0\} \ , \\
\Delta^{++}_\uparrow &=& \{2, 1, 0, 0\} \ .
\end{eqnarray}
Strange quarks are also straightforwardly added
\begin{eqnarray}
&& \Sigma^+_\uparrow = \sqrt{\frac23} \{2, 0, 0, 0\} s_\downarrow
- \frac{1}{\sqrt3} \{1, 1, 0, 0\} s_\uparrow \ , \\
&& \Lambda^0_\uparrow = \Big( \frac{1}{\sqrt2} \{1, 0, 0, 1\}
- \frac{1}{\sqrt2} \{0, 1, 1, 0\} \Big) s_\uparrow \ .
\end{eqnarray}
We could have equally well written these states in terms of the ``6n-symbol''
introduced above,
but we gave them here in a form which is not symmetrized under the exchange of
the $s$ quark with the $u,d$ quarks. This is appropriate in broken SU(3)
and for hadrons containing one heavy quark, with the replacement $s\to Q$.

Exotic states containing both
quarks and antiquarks can also be constructed. We consider here only
positive parity pentaquark-type states, containing $N_c+1$ quarks and one
antiquark. We take the $N_c+1$ quarks to
contain only u,d quarks in a spin-flavor symmetric state, as in Ref.\cite{JM3},
while the antiquark can be a strange or heavy quark.
Representative states with $I=0,1$ can be chosen as
\begin{eqnarray}
&& \Theta^+_\uparrow = \\
&& \quad \left( \frac{1}{\sqrt3} \{2,0,0,2\}
- \frac{1}{\sqrt3} \{1,1,1,1\} +
\frac{1}{\sqrt3} \{0,2,2,0\}\right) \bar s_\uparrow \nn \\
&& \Theta^{++}_{1\uparrow}(I_3=+1) =  \\
&& \qquad \left( \frac{1}{\sqrt2} \{3,0,0,1\}
- \frac{1}{\sqrt6} \{2,1,1,0\} \right) \bar s_\downarrow  \nn \\
&& \qquad - \left( \frac{1}{\sqrt6} \{2,1,0,1\}
- \frac{1}{\sqrt6} \{1,2,1,0\} \right) \bar s_\uparrow\,. \nn
\end{eqnarray}

Next we consider the action of quark operators on these states.
Any such operator can be constructed from one-body annihilation $q_i$
and creation $q_i^\dagger$ operators, where $i=1-6$ denotes one of the
basis states in Eq.~(\ref{basis}). Their action is given explicitly as
\begin{eqnarray}\label{rules}
&& q_i \{ \cdots, n_i, \cdots \} =
\sqrt{n_i} \{ \cdots, n_i-1, \cdots \} \ , \\
&& q_i^\dagger \{ \cdots, n_i, \cdots \} =
\sqrt{n_i+1} \{ \cdots, n_i+1, \cdots \} \,.
\end{eqnarray}
with all occupation numbers $n_{j\neq i}$ unchanged. They satisfy the
usual commutation relations for bosonic operators $[q_i, q^\dagger_j]=
\delta_{ij}$.

As an example of their application, consider
the isospin lowering and raising operators. Their action can be
obtained by first writing them in terms of quark
operators
\begin{eqnarray}
I_- = d_\uparrow^\dagger u_\uparrow + d_\downarrow^\dagger u_\downarrow \,,\quad
I_+ = u_\uparrow^\dagger d_\uparrow + u_\downarrow^\dagger d_\downarrow
\end{eqnarray}
followed by the application of the rules Eqs.~(\ref{rules}). One finds
\begin{eqnarray}
&& I_- \{ n_1, n_2, n_3, n_4 \} = \\
&& \qquad\,\,\, \sqrt{n_1 (n_3+1)} \{ n_1-1, n_2, n_3+1, n_4 \}\nn \\
&& \qquad + \sqrt{n_2 (n_4+1)} \{ n_1, n_2-1, n_3, n_4+1 \} \ , \nn \\
&& I_+ \{ n_1, n_2, n_3, n_4 \} = \\
&& \qquad\,\,\, \sqrt{n_3 (n_1+1)} \{ n_1+1, n_2, n_3-1, n_4 \}\nn \\
&& \qquad + \sqrt{n_4 (n_2+1)} \{ n_1, n_2+1, n_3, n_4-1 \} \nn \,.
\end{eqnarray}
They are useful for obtaining the entire isospin multiplets from the
representative states listed above.
\vspace{0.5cm}

3. \textit{Matrix elements.}
In broken SU(3), the SU(6) generators in Eq.~(\ref{generators})
can be decomposed into generators of the subgroup Eq.~(\ref{decomp})
plus operators mediating transitions between sectors of different
$n_s$. They can be chosen as
\begin{eqnarray}\label{bblocks}
&& J^i  \ \ , \ \ I^a = T^a \ \ , \ \ G^{ia} = G^{ia}  \qquad (i,a = 1...3)  \ , \nonumber \\
&& \tilde t^\alpha =  q^{\dagger\alpha} s \ \ , \ \
t_\alpha = s^\dagger q_\alpha \, \ \ \ \ \ \ \ \ \ \  \quad (\alpha = \pm 1/2) \ , \nonumber \\
&& \tilde Y^{i\alpha} =  q^{\dagger\alpha} \frac{\sigma^i}{2} s
\ \ , \ \
Y^i_\alpha = s^\dagger \frac{\sigma^i}{2} q_\alpha  \ , \\
&& J_s^i = s^\dagger \frac{\sigma^i}{2}  s \ , \ N_s = s^\dagger s \,. \nonumber
\end{eqnarray}
where $q^{\dagger\alpha} = (u^\dagger , d^\dagger)^\alpha$ and
$q_{\alpha} = (u , d)_\alpha$. The adjoint of a spherical tensor
operator $O^{j,m}$ is defined in terms of its components as
$O^{\dagger j,m} \equiv (O^{j,-m})^\dagger (-1)^{j-m} =
(O^j_{m})^\dagger $, where indices are raised and lowered by
contracting with the metric tensor \cite{edmonds}
{\small
\begin{eqnarray}
 O^j_m = \sum_{m'} \left(
\begin{array}{c}
 j  \\
 m \ m'
\end{array}
\right) O^{j m'}
\ , \
O^{j m} =
\sum_{m'} \left(
\begin{array}{c}
 j  \\
 m' \ m
\end{array}
\right) O^j_{m'}
\end{eqnarray}
}
defined as
\begin{eqnarray}
\left(
\begin{array}{c}
 j  \\
 m \ m'
\end{array}
\right)  = (-1)^{j-m} \delta_{m,-m'} \ .
\end{eqnarray}
The most general $n$-body operator can be constructed from the
building blocks shown in Eq.~(\ref{bblocks}).

The matrix elements of the strangeness conserving operators
$I^a, G^{ia}, N_s, J_s^i$ can be obtained using well-known SU(2)
methods \cite{KaPa,Soldate}.
The matrix element of $G^{ia}$ on states with
$N_{ud} = N_c-n_s$ up and down quarks are given by
\begin{eqnarray}
&& \langle I' I_3' J_3'|G^{i a}|I I_3 J_3 \rangle = \\
&& \quad \frac{1}{2I'+1} X^{(N_{ud})}(I',I)
\left(
\begin{array}{cc|c}
I   &  1 &  I'\\
I_3 &  a & I'_3
\end{array}
\right)
\left(
\begin{array}{cc|c}
J &  1 &  J'\\
J_3 & i & J'_3
\end{array}
\right)
\nonumber
\end{eqnarray}
with $X^{(N_{ud})}(I',I) = X^{(N_{ud})}(I,I')$, given explicitly as
\begin{eqnarray}
&& X^{(N_{ud})}(I',I) =  \sqrt{(2I'+1)(2I+1)} \\
&& \quad \times \sqrt{[(N_{ud}+2)^2 - (I'-I)^2 (I'+I+1)^2]} \ . \nonumber
\end{eqnarray}
For the matrix element of $G^{ia}$ on general states $|JIn_s\rangle$ containing
$N_{ud}$ $u$,$d$ quarks and $n_s$ strange quarks we obtain
\begin{eqnarray}
&& \langle I' I_3' , J'J_3' ;n_s| G^{ia}| I I_3 , J J_3 ;n_s\rangle = \\
&& \quad
\left(
\begin{array}{cc|c}
I &  1 &  I'\\
I_3 &  a & I'_3
\end{array}
\right)
\left(
\begin{array}{cc|c}
J &  1 &  J'\\
J_3 & i & J'_3
\end{array}
\right)
X(I'J',IJK) \ ,
\nonumber
\end{eqnarray}
with
\begin{eqnarray}
&& X(I'J',IJK) = \sqrt{\frac{2J+1}{2I'+1}} X^{(N_{ud})}(I',I) \\
&& \quad \times
(-)^{J+K+I'}
\left\{
\begin{array}{ccc}
1 & I & I' \\
K & J' & J \\
\end{array}
\right\} \ . \nn
\end{eqnarray}
This matrix element has a $1/N_c$ expansion of the form
\begin{eqnarray}
&& X(I'J',IJK) = N_c X_0(I'J', IJK)\\
&& \qquad  + X_1(I'J', IJK) + \cdots \nn
\end{eqnarray}
with the first two terms $X_{0,1}$ in agreement with the model-independent
prediction following from the contracted symmetry \cite{DJM}.

In the following we compute also the matrix elements of
the two strangeness lowering operators $t^\alpha, Y^{i\alpha}$  (with raised indices),
and the two strangeness raising operators $\tilde t^\alpha , \tilde Y^{i\alpha}$.

We define the reduced matrix elements as
\begin{eqnarray}\label{Ttildedef}
&& \langle I' I_3' , J'J_3' ;n_s-1|\tilde Y^{i\alpha }| I I_3 , J J_3 ;n_s\rangle
=\\
&& \qquad \qquad
\left(
\begin{array}{cc|c}
I &  \frac12 &  I'\\
I_3 &  \alpha & I'_3
\end{array}
\right)
\left(
\begin{array}{cc|c}
J &  1 &  J'\\
J_3 & i & J'_3
\end{array}
\right)
 \tilde Y(I'J'K',IJK)
\nonumber
\end{eqnarray}
and  similarly for $Y^{i\alpha }$, in terms of $Y(I'J'K',IJK)$.
The reduced matrix elements of $t^\alpha$ and $\tilde t^\alpha$ are
defined as
\begin{eqnarray}
&& \langle I' I_3' , J'J_3' ;n_s-1|\tilde t^{\alpha }| I I_3 , J J_3 ;n_s\rangle
=\\
&& \qquad \qquad  \delta_{JJ'} \delta_{J_3 J_3'}
\left(
\begin{array}{cc|c}
I &  \frac12 &  I'\\
I_3 &  \alpha & I'_3
\end{array}
\right) \tilde t(I'K',IJK) \nn
\end{eqnarray}
and similarly for $t(I'K',IJK)$.

\begin{table}
\caption{\label{table2ns1} Reduced matrix elements  $\tilde Y$
and $\tilde t$ for $(sq^{N_c-1}) \to (q^{N_c})$ transitions. }
\begin{ruledtabular}
\begin{tabular}{r|l|cc}
Transition & $(I'J',IJ)$ & $\tilde Y(I'J'K',IJK)$ & $\tilde t(I'K',IJK)$ \\
\hline $\Lambda \to N\bar K$ & $(\frac12\frac12, 0\frac12 )$
   & $\frac{\sqrt3}{2}\sqrt{N_c+3}$   & $\frac12\sqrt{N_c+3}$ \\
\hline $\Sigma \to N\bar K$ & $(\frac12\frac12, 1\frac12 )$
   & $-\frac12\sqrt{N_c-1}$  & $\frac{\sqrt3}{2} \sqrt{N_c-1}$ \\
$\to \Delta \bar K$ & $(\frac32\frac32, 1\frac12 )$
   & $-\frac{1}{\sqrt2}\sqrt{N_c+5}$ & $-$ \\
\hline
$\Sigma^* \to N\bar K$ & $(\frac12\frac12, 1\frac32)$  & $\sqrt2\sqrt{N_c-1}$ &  $-$ \\
$\to \Delta \bar K$ & $(\frac32\frac32, 1\frac32)$  & $\frac12
\sqrt\frac52 \sqrt{N_c+5}$ &
 $\frac12 \sqrt\frac32 \sqrt{N_c+5}$\\
\end{tabular}
\end{ruledtabular}
\end{table}

\begin{table}
\caption{\label{table2ns2} Reduced matrix elements  $\tilde Y$
and $\tilde t$ for $(ssq^{N_c-2}) \to (sq^{N_c-1})$ transitions. }
\begin{ruledtabular}
\begin{tabular}{r|l|cc}
Transition & $(I'J',IJ)$ & $\tilde Y(I'J'K',IJK)$ & $\tilde t(I'K',IJK)$ \\
\hline
$\Xi \to \Sigma\bar K$ & $( 1\frac12 , \frac12\frac12)$
 & $\frac{5}{3 \sqrt{2}} \sqrt{N_c+3}$ &$\frac{1}{\sqrt{6}} \sqrt{N_c+3}$ \\
$\to \Sigma^* \bar K$ & $( 1\frac32, \frac12\frac12)$
& $-\frac{\sqrt{2}}{3} \sqrt{N_c+3}$ & $-$ \\
$\to \Lambda\bar K$ & $( 0\frac12, \frac12\frac12)$
 &  $\frac{1}{\sqrt{2}} \sqrt{N_c-1}$ & $\sqrt{\frac{3}{2}} \sqrt{N_c-1}$\\
\hline
$\Xi^* \to \Sigma\bar K$ & $( 1\frac12, \frac12\frac32)$
& $-\frac{2}{3} \sqrt{N_c+3}$ & $-$\\
$\to \Sigma^* \bar K$ & $( 1\frac32, \frac12\frac32 )$
& $\frac{\sqrt{10}}{3} \sqrt{N_c+3}$  & $ \sqrt{\frac23} \sqrt{N_c+3}$\\
$\to \Lambda\bar K $ & $( 0\frac12, \frac12\frac32)$ &  $2\sqrt{N_c-1}$ & $-$ \\
\end{tabular}
\end{ruledtabular}
\end{table}

The action of the strangeness changing operators on quark states
can be obtained straightforwardly using the rules  Eq.~(\ref{rules}). Typical
relations are
{\small
\begin{eqnarray}
&& (u^\dagger \sigma^3 s) \{n_1,n_2,n_3,n_4\} s_\downarrow =
- \sqrt{n_2+1 } \{ n_1, n_2+1, n_3, n_4\}  \ , \nn \\
&& (u^\dagger s) \{n_1,n_2,n_3,n_4\} s_\uparrow= \sqrt{n_1+1}
\{ n_1+1, n_2, n_3, n_4\}  \ .
\end{eqnarray}
}
Repeated application of the $s$ creation operators leads to states
containing multiple strange quarks. Typical matrix elements that
appear for example in the computation of $\Xi \to \Sigma$ matrix elements are
(in the notation of Eq.~(\ref{nons}) for the nonstrange states)
\begin{eqnarray}
&\langle&\!\! 1 1 1 ; N_c-1 |  u^\dagger_\uparrow  | \frac12 \frac12 \frac12 ; N_c - 2 \;\; \rangle = \nonumber \\
&=& \sum_{i=-\frac{N_c-5}{4}}^\frac{N_c-1}{4}\left(
\begin{array}{cc|c}
\frac{N_c+1}{4} & \frac{N_c-3}{4} & 1 \\
i + \frac12 & \frac12 - i  & 1
\end{array}
\right) \left(
\begin{array}{cc|c}
\frac{N_c-1}{4} & \frac{N_c-3}{4} & \frac{1}{2} \\
i  & \frac12 - i  & \frac{1}{2}
\end{array}
\right)   \nonumber \\
&\times& \sqrt{\frac{N_c+3}{4} + i} \  =  \ \sqrt{\frac{N_c}3 + 1} \ .
\end{eqnarray}
We show in Tables I, II the results for the strangeness raising transitions
$\tilde Y(I'J'K',IJK)$ and $\tilde t(I'K',IJK)$ for all ground state baryons.

\begin{table}
\caption{\label{table2alt} The reduced matrix elements $Y$
and $t$ for $\Theta\to NK, \Delta K$ transitions.
}
\begin{ruledtabular}
\begin{tabular}{r|l|cc}
Transition & $(I'J',IJ)$ & $Y(I'J'K',IJK)$ & $t(I'K',IJK)$ \\
\hline
$\Theta_0(\frac12) \to NK$ & $( \frac12\frac12, 0\frac12)$  &
                         $-\frac{\sqrt3}{2}\sqrt{N_c+1}$ & $\frac12 \sqrt{N_c+1}$ \\
\hline
$\Theta_1(\frac12) \to NK$ & $(\frac12\frac12, 1\frac12)$  &
                             $\frac12\sqrt{N_c+5}$ &
                             $\frac{\sqrt3}{2}\sqrt{N_c+5}$ \\
$\to \Delta K$ & $(\frac32\frac32, 1\frac12)$  &
                             $\frac{1}{\sqrt2}\sqrt{N_c-1}$ & $-$ \\
\hline
$\Theta_1(\frac32) \to NK$ & $(\frac12\frac12, 1\frac32)$  &
                             $-\sqrt2\sqrt{N_c+5}$ & $-$ \\
$\to \Delta K$ & $(\frac32\frac32, 1\frac32)$  &
                 $-\frac12\sqrt{\frac52}\sqrt{N_c-1}$ &
                 $\frac12{\sqrt\frac32} \sqrt{N_c-1} $ \\
\hline
 $\Theta_2(\frac32)\to \Delta K$ & $(\frac32\frac32 ,
2\frac32)$  &
                 $\frac12\sqrt{\frac32}\sqrt{N_c+7}$ &
                 $\frac12 \sqrt{\frac52}\sqrt{N_c+7}$ \\
\hline
$\Theta_2(\frac52)\to \Delta K$ & $(\frac32\frac32,2\frac52)$  &
                 $-\sqrt{\frac32}\sqrt{N_c+7}$ & $-$ \\
\end{tabular}
\end{ruledtabular}
\end{table}

In Table III we give also the results for
the kaon decays of pentaquarks with symmetric spin-flavor wave function.
The computation of matrix elements of operators containing
antiquarks requires some care.
The antiquark spin doublet $\bar s$ has components $\bar s^\beta =
(-\bar s_\downarrow, \bar s_\uparrow)^\beta$.
 This can be used to
express the operators $Y^i_\alpha, t_\alpha$ in terms of quark and
antiquark one-body operators as
\begin{eqnarray}
&& Y^3_\alpha = s^\dagger\frac{\sigma^3}{2} q_\alpha  - \bar s\frac{\sigma^3}{2} q_\alpha =  \\
&& \qquad \qquad
\frac12 (s_\uparrow^\dagger q_{\alpha\uparrow} - s_\downarrow^\dagger q_{\alpha\downarrow}
+ \bar s_\downarrow q_{\alpha\uparrow} + \bar s_\uparrow q_{\alpha\downarrow} ) \ , \nn \\
&& t_\alpha = s^\dagger q_\alpha - \bar s q_\alpha =  \\
&& \qquad \qquad s_\uparrow^\dagger q_{\alpha\uparrow} + s_\downarrow^\dagger q_{\alpha\downarrow}
+ \bar s_\downarrow q_{\alpha\uparrow} - \bar s_\uparrow q_{\alpha\downarrow} \ . \nn
\end{eqnarray}

For example, the action of a typical strangeness changing operator on pentaquark
states containing a $\bar s$ quark is given by
\begin{eqnarray}
\bar s \sigma^3 u \{n_1, n_2, n_3, n_4 \} \bar s_\uparrow =
  -\sqrt{n_2 } \{n_1, n_2-1, n_3, n_4 \}.
\end{eqnarray}
The results for the reduced matrix elements $Y,\tilde Y$ and $t,\tilde t $ take a simpler form
in the large $N_c$ limit, and can be given in analytical form. Expanding
these operators as
\begin{eqnarray}
&& Y = \sqrt{N_c} \Big( Y_0  + \frac1{N_c} Y_1 + \cdots  \Big)
\end{eqnarray}
and similarly for $t$ and the strangeness raising operators,
the reduced matrix elements at leading order are given by (with $[I]\equiv 2I+1$)
\begin{eqnarray}\label{Ksol}
&& \tilde Y_0(I'J'K',IJK) = -Y_0(I'J'K',IJK) \nn \\
&& =
c(K,K') \sqrt{[I][J]}
\left\{
\begin{array}{ccc}
\frac12 & 1 & \frac12 \\
I & J & K \\
I' & J' & K' \\
\end{array}
\right\}
\end{eqnarray}
and
\begin{eqnarray}\label{tsol}
 &&\tilde t_0(I'K',IJK) = t_0(I'K',IJK) = \nn \\
 && = d(K,K') (-)^{J+I+K}\sqrt{[I]}
\left\{
\begin{array}{ccc}
K & K' & \frac12 \\
I' & I & J \\
\end{array}
\right\}\,.
\end{eqnarray}

The result Eq.~(\ref{Ksol}) was found in Ref.~\cite{DJM} using the method of the
induced representations for the contracted symmetry; the result Eq.~(\ref{tsol}) is new.
From Tables I, II and III we find $c(1,\frac12) = 3\sqrt{2}, c(\frac12, 0)=\sqrt{6}$
and $d(1,\frac12)=\sqrt{3}, d(\frac12,0)=1$.
\vspace{0.5cm}

4. \textit{Discussion and extensions.}  The applications
discussed so far were limited to symmetric spin-flavor states. The methods
of this paper can be extended also to mixed-symmetric states, which are
relevant for the application of the $1/N_c$ expansion to orbitally
excited baryons in the $\mathbf{70}^-$ \cite{Goity,SU3,PS1}, and to negative parity
exotic states \cite{Meg,PS3}.

The corresponding decomposition in broken SU(3) of a
state transforming in the mixed-symmetric ${\cal MS}_{N_c}$ representation of
SU(6) is more complicated than that for the symmetric state Eq.~(\ref{SNcdecomp}).
Keeping only terms with $n_s = 0,1$, this has the form
\begin{eqnarray}\label{MSNcdecomp}
{\cal MS}_{N_c} &=& (MS_{N_c}, {\mathbf 1}, 0)  \\
&\oplus &\,\, (MS_{N_c-1}, {\mathbf 2}, 1) \oplus
(S_{N_c-1}, {\mathbf 2}, 1) \oplus
\cdots \nn
\end{eqnarray}
where the two terms with one strange quark have the wave function of the
nonstrange system in a $MS$ and $S$ representation of SU(4), respectively.
The states in $(MS_{N_c-1}, {\mathbf 2}, 1)$ and $(S_{N_c-1}, {\mathbf 2}, 1)$ that
fall into ${\cal MS}_{N_c}$ can be constructed
by requiring that they are eigenstates of the quadratic SU(6) Casimir $C_2 = \sum_A \Lambda^A \Lambda^A$
with the eigenvalue of the ${\cal MS}_{N_c}$ representation.

The $MS$ states of $N_c-1$ nonstrange quarks can be represented in the usual way \cite{Goity,PY1,SU3}
as tensor products of a spin-flavor symmetric ``core'' of $N_c-2$ quarks with one ``excited'' quark.
This can be accomodated in our formalism in a similar way as we treated the strange quark.
New annihilation and creation operators acting on this ``excited'' one-body state have to be introduced,
from which transition operators can be constructed in the usual way.

In conclusion, we presented in this paper a new general method which simplifies computations
of matrix elements for ordinary and exotic baryon states, containing both light and strange
or heavy quarks, for arbitrary number of colors $N_c$.
The main advantage of the method is that only SU(2) Clebsch-Gordan coefficients need to
be used at any stage, and closed form results can be obtained for all matrix elements.

\vspace{0.5cm}

\begin{acknowledgments}
This work was supported by the DOE under cooperative research agreement
DOE-FC02-94ER40818 (D.P.). The work of C.S. was supported in part by Fundaci\'on Antorchas.
\end{acknowledgments}

\end{document}